\documentclass[twocolumn,twoside,slac]{revtex4}
\usepackage{graphicx}
\usepackage{fancyhdr}
\newcommand{\epp}{\epsilon^\prime/\epsilon}
\usepackage{relsize}
\usepackage{xspace}
\def\babar{\mbox{\slshape B\kern-0.1em{\smaller A}\kern-0.1em
    B\kern-0.1em{\smaller A\kern-0.2em R}}\xspace}
\def\Bbar  {\kern 0.18em\overline{\kern -0.18em B}{}\xspace}
\def\CP                 {\ensuremath{C\!P}\xspace}
\def\Bz    {\ensuremath{B^0}\xspace}
\def\B     {\ensuremath{B}\xspace}
\def\stwob{\ensuremath{\sin\! 2 \beta   }\xspace}

\begin{document}

%Title of paper
\title{Blind Analysis in Particle Physics}

% Repeat the \author .. \affiliation  etc. as needed
%
% \affiliation command applies to all authors since the last
% \affiliation command. The \affiliation command should follow the
% other information

\author{Aaron Roodman}
\affiliation{Stanford Linear Accelerator Center, Stanford, CA 94025, USA}

\begin{abstract}

A review of the blind analysis technique, as used in particle physics
measurements, is presented.  The history of blind analyses in physics
is briefly discussed.  Next the dangers of \emph{
  experimenter's bias} and the advantages of a blind analysis are
described. Three distinct kinds of blind analysis in particle physics
are presented in detail.  Finally, the \babar collaboration's experience
with the blind analysis technique is discussed.
\end{abstract}

%\maketitle must follow title, authors, abstract
\maketitle

\thispagestyle{fancy}

% body of paper here - Use proper section commands
% References should be done using the \cite, \ref, and \label commands
% Put \label in argument of \section for cross-referencing
%\section{\label{}}

\section{Introduction}

A \emph{blind analysis} is a measurement which is performed without looking
at the answer.  Blind analyses are the optimal way to reduce or
eliminate \emph{experimenter's bias}, the unintended biasing of a
result in a particular direction.

In bio-medical research the double-blind randomized clinical trial is
the standard way to avoid bias. In such experiments both patients and clinicians are
blind to the individual assignments of treatments to patients, and that
assignment is made randomly.  A double-blind
randomized trial was first used in 1948 by Hill in a study of
antibiotic treatments for tuberculosis\cite{Doll}.  Amazingly, the concept of a
double-blind trial dates back to at least 1662, when John Baptista van
Helmont made the following challenge\cite{Doll}:
\begin{quote}
  Let us take out of the hospitals,... 200, or 500
  poor People, that have Fevers, Pleurisies, etc. Let us
  divide them into half, let us cast lots, that one half of
  them may fall to my share, and the other to yours; I will cure them
  without blood-letting and sensible evacuation... We
  shall see how many funerals both of us shall have. But
  let the reward of the contention or wager, be 300
  florens, deposited on both sides...
\end{quote}

A notable early use of a blind analysis in physics was in a
measurement of the $e/m$ of the electron, by
Dunnington~\cite{Dunnington}.  In this measurement, the $e/m$ was
proportional to the angle between the electron source and the
detector. Dunnington asked his machinist to arbitrarily choose an angle around
$340^{o}$.  Only when the analysis was complete, and Dunnington was ready
to publish a result, did he accurately measure the \emph{hidden}
angle.

\section{Experimenter's Bias}

Experimenter's bias is defined as the unintended influence on a
measurement towards prior results or theoretical expectations. Next,
we consider some of the ways in which an unintended bias could be
present in a measurement.

\begin{figure}
  \includegraphics[width=0.95\columnwidth]{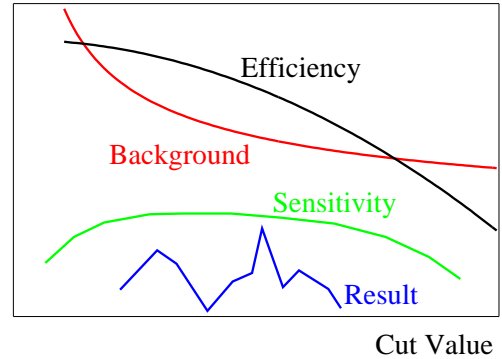}%
  \caption{\label{CutValue} This cartoon illustrates how a result may
    vary, statistically, for different arbitrary choices of a cut.}
\end{figure}

One scenario involves the choice of experimental selection
requirements, or cuts.  Often, a measurement may be equally well done,
in terms of sensitivity or uncertainties, with a range of values for a
particular selection cut, and the exact cut value used may be chosen
arbitrarily.  This is illustrated in the cartoon in Fig~\ref{CutValue},
where there is a plateau in the sensitivity, and the cut value could
be chosen anywhere in that plateau. However, the value of the result
may vary, typically within the statistical uncertainty, depending on
the exact value of cut chosen.  If the cuts are set with knowledge
of how that choice affects the results, experimenter's bias could
occur.  In this case, the size of the bias could be on the order of
the statistical uncertainty.

Another, less subtle, scenario involves measurements of small signals,
such as the search for rare processes or decays. Here experimenter's
bias could occur if the event selection is determined with prior
knowledge of the effect of that selection on the data. One danger is
that the selection cuts can be tuned to remove a few extra background-like
events, yielding a result biased to lower limits.  Another danger is
that the cuts can be tuned to improve the statistical significance
of a small signal.  

In general, experimenter's bias may occur if obtaining the
\emph{correct} result is the standard used to evaluate the quality of
the measurement.  The validity of a measurement may be checked in a
number of ways, such as internal consistency, stability under
variations of cuts, data samples or procedures, and comparisons
between data and simulation.  The numerical result, and how well it
agrees with prior measurements or the Standard Model, contains no real
information about the internal correctness of the measurement.  If
such agreement is used to justify the completion of the measurement,
then possible remaining problems could go unnoticed, and an
experimenter's bias occur.

\begin{figure}
  \includegraphics[width=0.95\columnwidth]{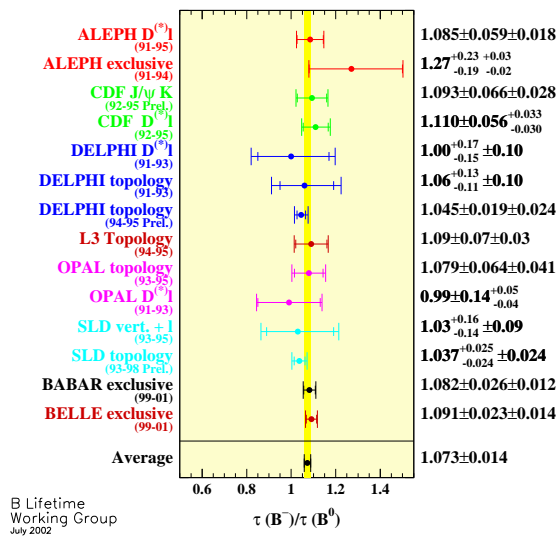}%
  \caption{\label{BLife} Summary of $B$ meson lifetime ratio measurements.
    The average has a $\chi^{2} = 4.5$ for 13 degrees of freedom.}
\end{figure}

Does experimenter's bias occur in particle physics measurements?
Consider the results on the ratio $B$ meson lifetimes shown in
Figure~\ref{BLife}.  The average has a $\chi^{2} = 4.5$ for 13 degrees
of freedom; a $\chi^{2}$ this small or smaller occurs only 1.5\% of the
time.  At this level, the good agreement between measurements is
suspicious, but for each individual result no negative conclusion
should be made.  Nonetheless, it can be argued that even the
possibility of a bias represents a problem.  The PDG\cite{pdg} has
compiled a number of measurements that have curious time-histories.
Likewise, while it is difficult to draw negative conclusions about a
single measurement, the overall impression is that experimenter's bias
does occur. Finally, there are numerous examples in particle physics
of small signals, on the edge of statistical significance, that turned
out to be artifacts.  Here too, experimenter's bias may have been present.

In all of these cases, the possibility of experimenter's bias is akin
to a systematic error.  Unlike more typical systematic effects, an
experimenter's bias cannot be numerically estimated.  Therefore, a
technique to reduce or eliminate this bias is needed.

\section{Blind Analysis}

A {Blind Analysis} is a measurement performed without looking at the
answer, and is the optimal way to avoid experimenter's bias.  A number
of different blind analysis techniques have been used in particle
physics in recent years.  Here, several of these techniques are
reviewed.  In each case, the type of blind analysis is well matched to
the measurement. 

\subsection{Hidden Signal Box}

The \emph{hidden signal box}  technique explicitly hides the signal
region until the analysis is completed.  This method is well suited
to searches for rare processes, when the signal region is known in
advance. Any events in the signal region, often in two variables, are
kept hidden until the analysis method, selection cuts, and background
estimates are fixed.  Only when the analysis is essentially complete is
the box opened, and an upper limit or observation made. 

\begin{figure}
  \includegraphics[width=0.95\columnwidth]{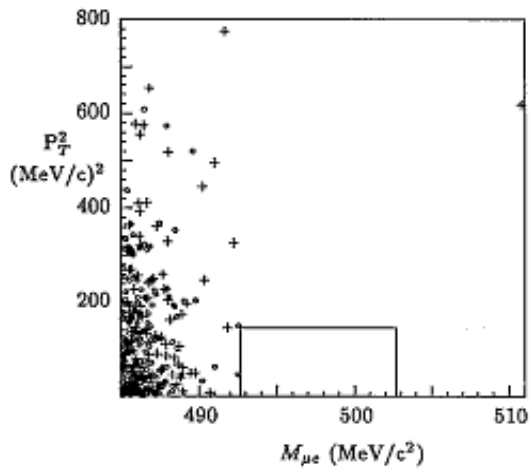}%
  \caption{\label{KLMuE} Hidden signal box from a search for the decay
  $K^0_{\scriptscriptstyle L} \rightarrow \mu^{\pm} e^{\mp}$ from
  Ref.~\cite{Molzon}. } 
\end{figure}

The \emph{hidden signal box} technique was used\footnote{This is the
  first use known to the author.} in a search for the rare decay
$K^0_{\scriptscriptstyle L} \rightarrow \mu^{\pm} e^{\mp}$. This
decay was not expected to occur in the Standard Model, and the single
event sensitivity of the experiment was one event in $10^{11}$
$K^0_{\scriptscriptstyle L}$ decays.  Any signal was expected inside
the box in $M_{\mu e}$ and $P^{2}_{T}$ shown in Figure~\ref{KLMuE};
the possible contents of this box were kept hidden until the analysis
was completed\cite{Molzon}.

The use of this method is now a standard method for rare decay
searches, when the signal region is known in advance.  One additional
subtlety lies in the size of the hidden box.  Generally, the box is
initially chosen to be somewhat larger than the signal region, so that
the final signal cuts may be chosen without bias as well.  Otherwise,
this technique is straightforward to apply.

\subsection{Hidden Answer}

For precision measurements of parameters, a different technique for
avoiding bias must be used.  In this case, \emph{hiding the answer} is
often the appropriate method.  The KTeV experiment used this technique
in its measurement of $\epsilon^\prime/\epsilon$.  The value of
$\epp$ was found in a fit to the data, and a small
value of order $10^{-4} - 10^{-3}$ was expected.  In this case, KTeV
inserted an unknown offset into its fitting program, so that the
result of the fit was the hidden value:
\begin{equation}\label{HideEqn}
  \epp \, ({\rm Hidden}) =  \left\{ 
    \begin{array}{c}1\\-1\end{array}
  \right\} 
  \times  \epp  + C 
\end{equation}
where C was a hidden random constant, and the choice of $1$ or $-1$
was also hidden and random.  The value of the hidden constant, $C$,
was made by a pseudo-random number generator with a reasonable
distribution and mean.  KTeV could determine its data samples,
analysis cuts, Monte-Carlo corrections, and fitting technique while
the result remained hidden, by looking only at $\epp \, ({\rm
  Hidden})$.  The use of the $1$ or $-1$ factor prevented KTeV from
knowing which direction the result moved as changes were made. In
practice, the result\cite{KTeV99} was unblinded only one week before
the value was announced. 

The hidden answer technique is well-suited to precise measurements of
a single quantity. The complete analysis, as well as the error
analysis, may proceed while blind to the result.  An additional
consideration is whether there are any distributions which will give
away the blinded result.  Often the exact value of the measurement is
not readily apparent from the relevant plots; in this case those plots
can be used without issue.

\subsection{Hidden Answer and Asymmetry}

\begin{figure}
  \includegraphics[width=0.95\columnwidth]{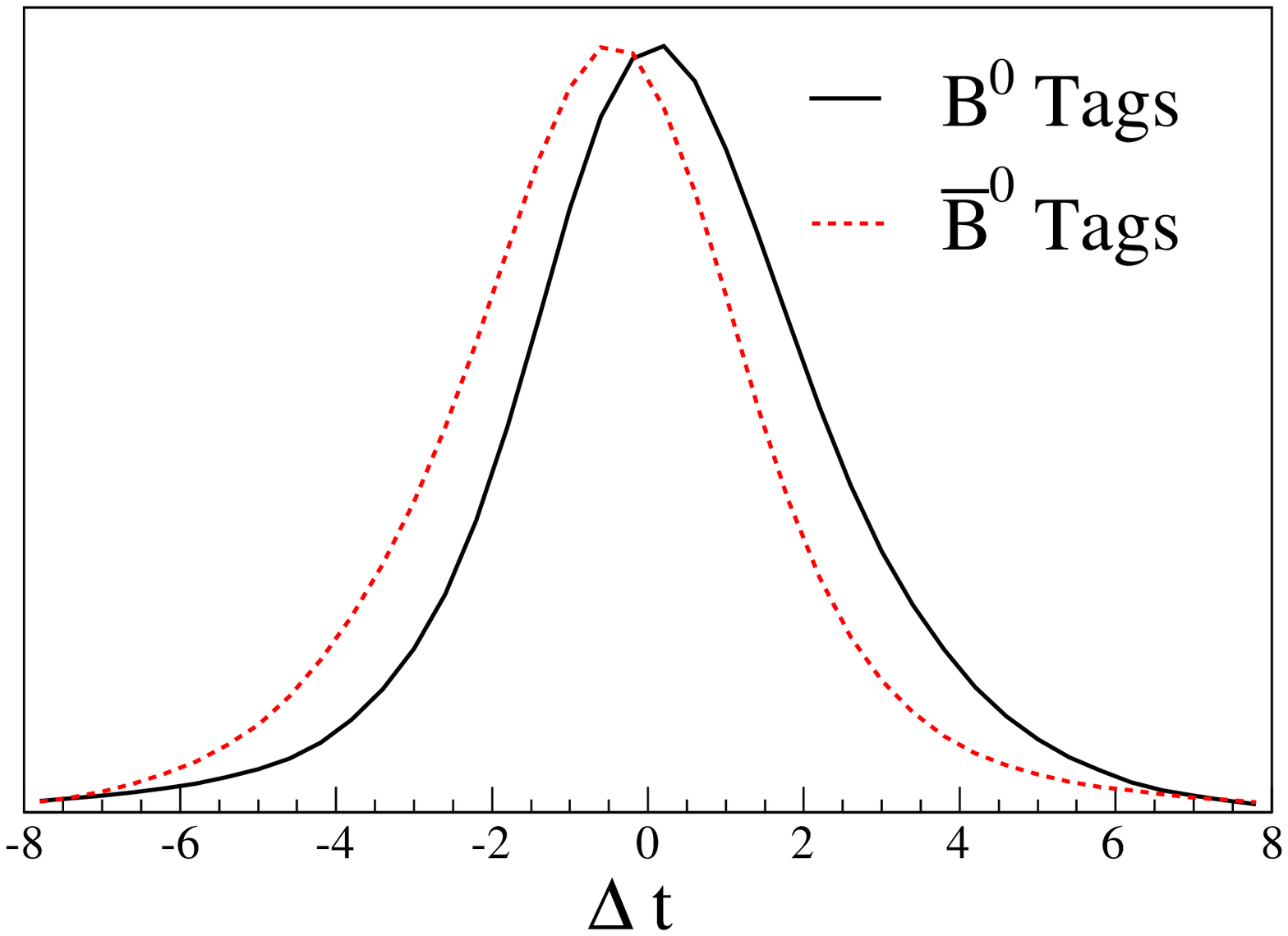}%

  \includegraphics[width=0.95\columnwidth]{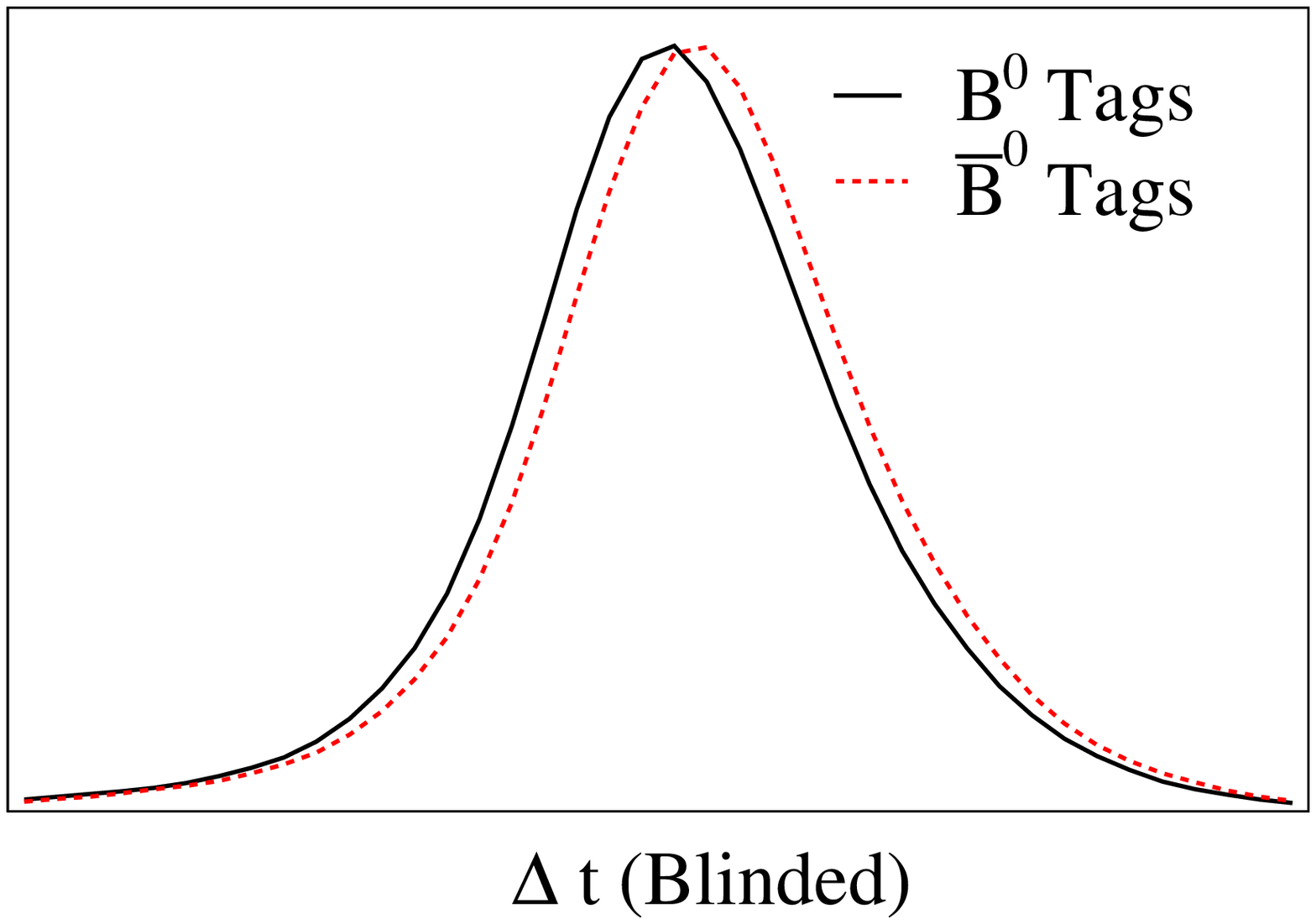}%
  \caption{\label{CPFig} The $\Delta t$ distributions for \B decays
    into \CP eigenstates, for $\stwob = 0.75$ with the \Bz flavor
    tagging and vertex resolution which are typical for the \babar
    experiment. a) The number of $B^{0}$ (solid line) and $\Bbar^{0}$
    (dashed line) 
    decays into $\CP$ eigenstates as a function of $\Delta t$. b) The
    $\Delta t_{\rm Blind}$ distributions for  $B^{0}$ (solid) and
    $\Bbar^{0}$ (dashed).
  }
\end{figure}

For certain measurements hiding the answer is not sufficient; it may
also be necessary to hide the visual aspect of the measurement.  One
example is an asymmetry measurement, such as the recent $\CP$-violation
measurement by \babar. In this case, the rough size and sign of the
asymmetry can be seen by looking at the $\Delta t$ distributions for
$B^{0}$ and $\Bbar^{0}$ decays into $\CP$ eigenstates, as shown in
Figure~\ref{CPFig}a.  Before $\CP$
violation had been established, and to avoid any chance of bias, a
blind analysis was developed to hide both the answer and the visual
asymmetry. 

In \babar's $\CP$-violation measurement the result, found from a fit to
the data, was hidden as in Equation~\ref{HideEqn}.  In addition, the
asymmetry itself was hidden by altering the $\Delta t$ distribution used to
display the data.\cite{BlindCP}  To hide the asymmetry the variable:
\begin{equation}\label{DtEgn}
  \Delta t \, ({\rm Blind}) =  \left\{ 
    \begin{array}{c}1\\-1\end{array}
  \right\} 
  \times  s_{\rm Tag} \times \Delta t  +  {\rm Offset} 
\end{equation}
was used to display the data.  The variable $ s_{\rm Tag}$ is equal to
$1$ or $-1$ for $B^{0}$ or $\Bbar^{0}$ flavor tags.  Since the
asymmetry is nearly equal and opposite for the different $B$ flavors,
we hid the asymmetry by flipping one of the distributions.  In
addition, the $\CP$-violation can be visualized by the asymmetry of
the individual $B^{0}$ and $\Bbar^{0}$ distributions.  In turn, this
was hidden by adding the hidden offset which has the effect of hiding
the $\Delta t = 0$ point.  The result is shown in Figure~\ref{CPFig}b,
where the amount of $\CP$-violation is no longer visible (the
remaining difference is due to charm lifetime effects). Also it is
worth noting that for a given data sample, due to statistical
fluctuations, the maximum of the distribution will not exactly
correspond to the $\Delta t = 0$ point, as in the smooth curves shown.

This blind analysis technique allowed \babar to use the $\Delta t_{\rm
  Blind}$ distribution to validate the analysis and explore possible
problems, while remaining blind to the presence of any
asymmetry. There was one additional restriction, that the result of
the fit could not be superimposed on the data, since the smooth fit
curve would effectively show the asymmetry.  Instead to assess the
agreement of the fit curve and the data, a distribution of just the
residuals was used.   In
practice, this added only a small complication to the measurement. 
However, after the second iteration of the measurement, it became
clear that the asymmetry would also remain blind if the only $\Delta
t$ distribution used was of the sum of $B^{0}$ and $\Bbar^{0}$
events, and that no additional checks were needed using the individual
$\Delta t$ distributions.  

\subsection{Other Blind Methods}

The kinds of measurements already discussed, such as rare searches and
precision measurements of physical parameters, are well suited to the
blind analysis technique.  
Other kinds of analyses are
difficult to adapt to the methods described.  For instance, branching
fraction measurements typically require the careful study of the
signal sample in both data and simulation, so it is not possible to
avoid knowing the number of signal events or the efficiency.  In this
case, other techniques may be considered.  One method is to fix
the analysis on a sub-sample of the data, and then used the identical
method on the full data sample.  One may argue about the correct
amount of data to use in the first stage, too little and backgrounds
or other complications may not be visible, too much and the technique
loses its motivating purpose.  Another method is to mix an unknown amount of
simulated data into the data sample, removing it only when the
analysis is complete. 

Another difficult example is the search for new particles, or
bump-hunting.  In this case, since the signal region is not known
a-priori, there is no one place to put a hidden signal box.  However,
such measurements may be the most vulnerable to the effects of
experimenter's bias. Certainly, there is some history of statistically
significant bumps that are later found to be artifacts.  The possibility of
using a blind analysis technique may depend on the understanding of
the relevant background.  If the background can be estimated
independently of the bump-hunting region, than the analysis and
selection cuts may be set independently of the search for bumps.  Here
again is a case in which the exact method used must be well matched to
the measurement in question.

\section{Conclusion}

The experience of the \babar collaboration in using blind analyses is
instructive. While the collaboration had initial reservations about
the blind analysis technique, it has now become a standard method for
\babar\cite{BabarBlind}. Often the blind analysis is a part of the
internal review of \babar results.  Results are presented and
reviewed, before they are unblinded, and changes are made while the
analysis is still blind.  Then when either a wider analysis group or
an internal review committee is satisfied with the measurement the
result is unblinded, ultimately to be published.  With several years of
data taking, and many results, \babar has successfully used blind
analyses.

% If you have acknowledgments, this puts in the proper section head.
\begin{acknowledgments}
Work supported by the U.S. Department of Energy under contract number DE-AC03-76SF00515.
\end{acknowledgments}

% Create the reference section using BibTeX:
%\bibliography{basename of .bib file}

\end{document}